\RequirePackage{lineno} 
\RequirePackage{fix-cm}

\documentclass[useAMS,usenatbib]{mn2e}
\usepackage{amsmath}            % Formeln
\usepackage{amssymb}            % Formeln
\usepackage{latexsym}
\usepackage{subfigure}
\usepackage{rotating}
\usepackage{graphicx}
\usepackage{dcolumn}% Align table columns on decimal point
\usepackage{bm}% bold math
\usepackage{natbib}
\usepackage{color,soul}

\newcommand{\eg}{e.g., }

\newcommand{\ie}{i.e., }

\newcommand{\sect}[1]{Section \ref{s:#1}}

\newcommand{\eqn}[1]{Eq.\ (\ref{e:#1})}

\newcommand{\fig}[1]{Fig.\ \ref{fig:#1}}

\newcommand{\figs}[2]{Figs.\ \ref{fig:#1}--\ref{fig:#2}}

\newcommand{\Fig}[1]{Figure \ref{fig:#1}}

\newcommand{\hide}[1]{} % to hide text sections

\title[Simulating regoliths in microgravity]
  {Simulating regoliths in microgravity}
\author[N. Murdoch et al.]
  {N.~Murdoch,$^{1,2,3}$\thanks{E-mail:
naomi.murdoch@isae.fr} 
  B.~Rozitis,$^2$ S.F.~Green,$^2$ P.~Michel,$^3$
  T-L.~de Lophem$^2$, and 
  \newauthor % starts a new line in the
               W.~Losert$^4$
             \\
 $^1$Institut Sup\'{e}rieur de l'A\'{e}ronautique et de l'Espace, 31055 Toulouse, France \\
  $^2$Department of Physical Sciences, The Open University, Milton Keynes, MK7 6AA, UK \\
  $^3$Laboratoire Lagrange, UMR7293, Universit\'{e} de Nice Sophia-Antipolis,  CNRS, Observatoire de la C\^{o}te d'Azur, 06300 Nice, France\\
  $^4$Institute for Physical Science and Technology, and Department of Physics, University of Maryland, USA }
\date{Accepted 2013 April 26. Received 2013 April 21; in original form 2013 February 20}

\pagerange{\pageref{firstpage}--\pageref{lastpage}} \pubyear{2013}

\def\LaTeX{L\kern-.36em\raise.3ex\hbox{a}\kern-.15em
    T\kern-.1667em\lower.7ex\hbox{E}\kern-.125emX}

\begin{document}
\graphicspath{{Figs/}{Figs/PDF}}

\label{firstpage}

\maketitle

\begin{abstract}

Despite their very low surface gravities, the surfaces of asteroids and comets are covered by granular materials - regolith - that can range from a fine dust to a gravel-like structure of varying depths. Understanding the dynamics of granular materials is, therefore, vital for the interpretation of the surface geology of these small bodies and is also critical for the design and/or operations of any device planned to interact with their surfaces.  We present the first measurements of transient weakening of granular material after shear reversal in microgravity as well as a summary of experimental results recently published in other journals, which may have important implications for small-body surfaces. Our results suggest that the force contact network within a granular material may be weaker in microgravity, although the influence of any change in the contact network is felt by the granular material over much larger distances. This could mean that small body surfaces are even more unstable than previously imagined. However, our results also indicate that the consequences of, \eg a meteorite impact or a spacecraft landing, may be very different depending on the impact angle and location, and depending on the prior history of the small body surface.

 \end{abstract}


%\linenumbers
\section{Introduction}\label{s:intro}

From solid planets to small bodies of our Solar System surface gravities vary by many orders of magnitude. Despite their very low surface gravities, the surfaces of small bodies are covered by granular materials that can range in size from a few microns (dust) or few hundreds of microns (sand) to a few centimetres or metres (gravels, pebbles, boulders). To date there have been three space missions sent to characterise an asteroid in detail: the NASA near-Earth asteroid rendez-vous (NEAR) mission which arrived at the near-Earth asteroid (433) Eros in 2000 \citep{cheng97}, the JAXA Hayabusa sample return mission which arrived at the near-Earth asteroid (25143) Itokawa in 2005, and the NASA Dawn mission to the main-belt asteroids (4) Vesta and (1) Ceres, which arrived at Vesta, the first of the two targets, in 2011 \citep{russell07}. There have also been other missions which have flown past asteroids. For example, in 2012 the Chinese spacecraft Chang'e-2 flew by asteroid (4179) Toutatis. The ESA Rosetta mission, whose primary target is Comet 67P/Churyumov-Gerasimenko, flew by asteroids (2867) Steins and (21) Lutetia in 2008 and 2010, respectively. The NEAR spacecraft flew by asteroid (253) Mathilde in 1997 before reaching Eros, the NASA Stardust spacecraft flew by (5535) Annefrank in 2002 en route to comet Wild 2, and the Galileo spacecraft flew by asteroids (951) Gaspra and (243) Ida in 1991 and 1993, respectively, on its way to Jupiter. 

The first two in-situ observations of asteroids (those of Eros and Itokawa) greatly increased our knowledge of asteroid surface properties whilst highlighting the complexities and variations in the surface environments.  Before the NEAR and Hayabusa space missions there was an on-going debate about whether asteroids, especially the smallest ones, were monolithic rocks with a bare surface or reaccumulated pieces of a larger asteroid that had been catastrophically distrupted by a collision \citep[\eg][]{chapman82,michel01}.  During these two space missions NEAR revealed a substantial regolith covering Eros \citep{robinson02}, and Hayabusa revealed Itokawa to be a rubble pile asteroid essentially made of regolith throughout \citep{fujiwara06}.   

In addition to finding each of these bodies to be regolith-covered, there is strong evidence that this regolith is very complex and active. In fact, it was due to the NEAR observations of Eros that the local gravity was first understood to be of importance to asteroid surface processes \citep{robinson02}. This was further emphasised when the first images were received from the Hayabusa probe.  Compared to Eros, Itokawa was found, astonishingly, to have entirely different structural and surface properties despite their similar taxonomic class. The reason for these different properties is not clearly understood, but perhaps this should not have been surprising; because of their size (mass) difference, if gravity is the discriminator, then Itokawa is expected to be as different from Eros, geologically, as Eros is from the Moon \citep{asphaug09}.

Understanding the physics of granular material in low-gravity environments is, therefore, important for the interpretation of spacecraft observations (images, spectral observations, or topography) but is also critical for the design and operations of any device that will interact will the small bodies' surface.  AstEx (ASTeroid Experiment), our parabolic flight experiment, was designed to study the mechanical response of granular material subject to shear forces in a microgravity environment using a Taylor-Couette shear cell. Here, in \sect{regdyn}, we bring to the attention of the planetary science community some experimental results published in other journals, which may have important implications for small-body surfaces. In addition, in \sect{expres}, we provide new experimental findings related to the reversal of shear direction of a granular material in a microgravity environment. Finally, in \sect{implications} we discuss the implications of these experimental findings for small body surfaces.

\section{Regolith dynamics in microgravity}\label{s:regdyn}

\subsection{Shear bands on asteroids}

Both fluids and granular materials react very differently to shear stresses. A fluid deforms uniformly whilst a granular material develops a shear band;  a narrow zone of large relative particle motion bounded with essentially rigid regions. In a shear cell set-up, almost all of the energy input into the granular system by the inner cylinder is dissipated by friction within this narrow region producing large velocity gradients. Shear bands mark areas of flow, material failure, and energy dissipation, and as such they are important in many geophysical processes. 

Impact phenomena \citep{paolicchi02, holsapple02}, tidal forces from planetary encounters \citep{bottke96}, and YORP spin up \citep{bottke06} could apply shear forces to the surface. Space agencies are planning sample return missions to near-Earth asteroids to bring back to Earth a pristine sample of an asteroidÕs surface \eg Osiris-Rex and Hayabusa-2 (NASA and JAXA missions, respectively) which have already been selected for flight and MarcoPolo-R, a proposed mission led by ESA. These missions may make use of a device which involves shearing the regolith material. 

Using a microgravity-modified Taylor-Couette shear cell under the conditions of parabolic flight microgravity, \cite{murdoch12d} show that the effect of constant shearing on a granular material in a direction perpendicular to the gravity field does not seem to be strongly influenced by gravity.  Specifically, it was found that, between the gravitational regimes of microgravity and 1 $g$, there is no difference in the width of the shear band nor is there a large difference in the magnitude of the angular (tangential) velocities within the shear band. The primary flow field exhibits shear banding, consistent with prior work in this geometry.  This means that shear bands can form on small body surfaces just as they do on Earth.

\subsection{Particle segregation in low-gravity}

In many granular flows where particles have different physical properties, particle segregation occurs. Granular avalanches are no different. For example, in an avalanche of a polydisperse granular material (\ie grains of many sizes), particles segregate according to their size with the smallest particles being nearest the bottom and the largest particles on the top. Since the small particles are concentrated at the bottom of the avalanche they are deposited first. The faster moving, large particles (the avalanche velocity is greatest at the surface) are transported to the front thus travelling furthest. There is evidence of a boulder-rich layer similar to segregated landslide deposits even on the tiny ($\sim$320 m) asteroid, Itokawa [\fig{MusesSeaA}] and particle segregation has also been observed on comet 103P/Hartley 2 [\fig{HARTLEY2}].

\begin{center}
%  [FIGURE 1 GOES HERE]
  [FIGURE \ref{fig:RegolithSeg} GOES HERE]
\end{center}

Segregation in a granular material also occurs during shaking; when a granular material is shaken the larger particles tend to move to the top. This resulting size segregation has been named the ``Brazil-nut effect" \citep{rosato87}. The Brazil-nut effect on asteroids may be activated during large enough but non-disruptive collisions, and may be responsible for the presence of large boulders on the surface of asteroids, such as Itokawa and Eros \citep{asphaug01, miyamoto07}.  If this is the case, the interiors of rubble-pile asteroids having experienced this kind of evolution are likely to be composed of smaller particles than those observed at the surface. This may also lead to some variations of macro-porosity with depth inside the asteroid. 

The dominant mechanism for segregation in granular avalanches, provided the density and size differences are not too large, is \emph{kinetic sieving}. As the grains avalanche downslope, there are fluctuations in the void space and the smaller particles are more likely to fall, under gravity, into gaps that open up beneath them because they are more likely to fit into the available space than the coarse grains. The fine particles, therefore, percolate towards the bottom of the flow, and force imbalances squeeze the large particles towards the surface \citep{gray06}.  It has also been suggested that kinetic sieving is responsible for the accumulation of larger particles at the ejecta flow front of single- and double-lobe impact craters on Mars \citep{baratoux05}. 

The mechanism driving the ``Brazil-nut'' segregation is still under debate. It has been suggested that the segregation results from the percolation of small particles in a similar fashion to the kinetic sieving mechanism, but here the local rearrangements are caused only by the vibrations \citep[\eg][]{rosato87, williams76}. However, other experimental results from \cite{knight93} have shown that vibration-induced size segregation arises from convective processes within the granular material and not always from local rearrangements. 

Granular convection is, in fact, a process often invoked by the community of small body scientists to interpret the surface geology of asteroids. \cite{miyamoto07} suggest that the arrangements of the gravels on Itokawa's surface are largely due to granular convective processes. It is suggested \citep{asphaug07} that piles of aligned boulders are formed where convection meets a boundary layer, perhaps analogous to miniature mountain belts. \cite{miyamoto07} have also proposed that apparent upstream sloping, or seas, on Itokawa could be caused by convective surfaces. However, since Itokawa has long lost any internal heat source capable of driving convection, the energy source would have to be a granular thermal input \citep{asphaug07}.  By this we mean any energy that leads to motion of grains. This energy could come, for example, from small impacts on the asteroid from micrometeorites.

One way of predicting the degree of segregation that will occur in a granular material due to kinetic sieving is to use the non-dimensional segregation number \citep{thornton05}. This number is formulated using the framework of mixture theory, which is often used to study porous media flow problems \citep{morland92}. However, the segregation rate is proportional to the gravitational acceleration and, therefore, this model predicts virtually no kinetic sieving based segregation in microgravity conditions.  Additionally, recent experiments \citep{murdoch12c} have demonstrated that granular convection is strongly dependent on the external gravitational acceleration. It is shown that convective-like flows in a granular material are close to zero in microgravity, and enhanced under high gravity conditions. \cite{murdoch12c} also suggest that gravity tunes the frictional particle-particle and particle-wall interactions, which have been proposed to drive the secondary flow.  Therefore, due to the strong dependence of kinetic sieving and granular convection on gravity, a weak gravitational acceleration will likely reduce the efficiency of particle size segregation.  Indeed, recent numerical simulations of the Brazil-nut effect have shown that the velocity at which a large intruder in a granular material rises is reduced as the external gravity decreases \citep{tancredi12}. 

As segregation in a granular material occurs in almost all granular flows where the particles have different physical properties it is likely that particle segregation does occur in the reduced-gravity environment found on asteroids' surfaces. However, all convective and particle segregation processes in a granular material on or near the surface of a small body may require much longer timescales than the same processes would require in the presence of a strong gravitational field. In addition, the reduction in inter-particle friction in low-gravity environments may have further consequences for the dynamics of regolith (\eg more fluidized behaviour than expected).

\section{Transient weakening of regolith in microgravity}\label{s:expres}

It has previously been shown that the flow of granular matter is strongly influenced by the network of direct contacts with neighbouring particles \citep{toiya04}. This contact network, in turn, is shaped by how the material evolved with time. When uniform shear or compression is applied a stronger contact network in the direction of forcing develops.  When the shear direction is reversed, or the direction of compression is changed, the material rearranges until it forms a new contact network that can best support the new direction of compression or shear \citep{toiya04}.  This is illustrated in the schematic in \fig{shear_reversal}; the force chains aligned to cause jamming in one direction are not suited to jam under the reverse driving. Therefore, time is required to re-form a force chain network subsequent to reversal of shear direction.   In a detailed analysis of the angular velocities of the particles \cite{toiya04} found that when sheared in the initial direction the system reaches steady state during which the outer regions of the shear cell do not experience any appreciable flow. If the driving is discontinued and then reapplied the system reaches the same steady state immediately. However, if the driving is stopped and then applied in the reverse direction, a transient sets in during which flow is evident both in the regions that were flowing and in the previously jammed regions. Average flowing velocities in all regions are faster initially and drop off with roughly the same time-scale in all regions. The implication is that regions that normally do not move under steady shear, move significantly during reversal of the shear direction \citep{toiya04}.  

\begin{center}
%    [FIGURE 2 GOES HERE]
    [FIGURE \ref{fig:shear_reversal} GOES HERE]
\end{center}

Studying the reversal of shear in a granular material in microgravity has the potential to shed light on different behaviours that are evident when granular material is sheared in different directions, such as the transient weakening found to occur on the ground. Additionally, the flow fields during the shear reversal experiments of \cite{toiya04} are accompanied by compaction due to gravity. It is not clear how the force chains would break and reform in the absence of a preferred guiding direction such as gravity.  

\subsection{Experimental set-up and procedures}

Our experiments use a Taylor-Couette geometry (\fig{TaylorCouette}).  There are two concentric cylinders. The outer cylinder is fixed and its inside surface is rough with a layer of particles, and the outer surface of the inner cylinder is also rough and rotated by a motor to generate shear strain.  The floor between the two cylinders is smooth and fixed in place. The gap between the two cylinders is filled, to a height of 100 mm, with spherical soda lime glass beads (grain diameter, $d$ $=$ 3 mm or 4 mm; density, $\rho$ $=$  2.55 g cm$^{-3}$) upon which the rotating inner cylinder applies shear stresses. A movable and transparent disk is used to confine the granular material during the microgravity phase of a parabola with an average force of 6.6 N (the force can vary from 0 to 13.2 N depending on the packing fraction of the granular material).  Grain sizes of 3 mm and 4 mm result in the gap between the two cylinders being $\ge$25 $d$ allowing particle dynamics both inside and outside the shear band to be studied.  These sizes of particles are also large enough not to cause problems between the moving parts that were designed with a tolerance of 1 mm, nor to be affected by electrostatic charging. Further details of our experimental design can be found in \cite{murdoch12d}.

\begin{center}
%  [FIGURE 3 GOES HERE]
  [FIGURE \ref{fig:TaylorCouette} GOES HERE]
\end{center}

In order to reduce the ambient gravitational acceleration during our experiments to levels close to those found at the surface of an asteroid we must find a way of achieving microgravity. Microgravity, the condition of relative near weightlessness, can only be achieved on Earth by putting an object in a state of free-fall. This is achieved using a parabolic flight. During each parabola of a parabolic flight there are three distinct phases: a $\sim$20 second $\sim$1.8 $g$ (where $g$ is the Earth's gravitational acceleration) injection phase as the plane accelerates upwards, a $\sim$22 second microgravity phase as the plane passes through the top of the parabola (during this period the pilot carefully adjusts the thrust of the aircraft to compensate for the air drag so that there is no lift), and finally, a $\sim$20 second $\sim$1.8 $g$ recovery phase as the plane pulls out of the parabola. During one flight there are $\sim$31 parabolas, and during one flight campaign there are normally 3 flights. This means that one flight campaign offers approximately 30 minutes of simulated microgravity.

The motor that drives the inner cylinder was started shortly after the microgravity phase begins for each parabola, and after $\sim$10 seconds the direction of rotation was reversed. High-speed cameras imaged the top and bottom layers of glass beads in the shear cell at $\sim$60 frames sec$^{-1}$ so that the particles did not move more than 1/10 $d$ between consecutive frames. In between the parabolas the shear cell is shaken by hand to attempt to reproduce the same initial bulk packing fraction while minimising possible memory effects from prior shear. 

After the flights, particle tracking was performed using an adaptation of a subpixel-accuracy particle detection and tracking algorithm \citep{crocker96}, which locates particles with an accuracy of approximately $1/10$ pixel. The particles are then separated, according to their radial position, into equal cylindrical rings (referred to as radial bins) reaching from the inner cylinder wall to the outer cylinder wall (\fig{RadialBins}).  

\begin{center}
 % [FIGURE 4 GOES HERE]
 [FIGURE \ref{fig:RadialBins} GOES HERE]
\end{center}

\subsection{Detecting transient weakening after shear reversal}

The mean normalised angular velocity, $V^*_{\theta}$, is given by, 

\begin{equation}
V^*_\theta =  \frac{V_{\theta}}{\omega}
\label{e:VstarTheta}
\end{equation}

\noindent where $V_{\theta}$ is the mean particle angular velocity and $\omega$ is the inner cylinder angular velocity.  By considering plots of the normalised angular velocity, $V^*_{\theta}$, as a function of inner cylinder displacement for each of the radial bins in a shear reversal experiment, we can examine the mean particle angular velocities before and after shear reversal. In \fig{detection} an example of a ground-based shear reversal experiment with 4 mm particles is given. The moment when shear reversal occurs is when the angular velocity of the inner (green) radial bin passes through zero.  Shown for clarity are both the unnormalised and normalised plots.  Just after the point of shear reversal, there is a transient period during which the average flowing velocities in the previously flowing regions are faster. These increased velocities drop off with roughly the same timescale in all regions.  This implies that transient weakening does occur on the top surface of our experiment on the ground.

\begin{center}
%  [FIGURE 5 GOES HERE]
  [FIGURE \ref{fig:detection} GOES HERE]
\end{center}

\Fig{ShearRevProfs_4mm0g} shows the normalised angular velocity as a function of inner cylinder displacement for an example shear reversal experiment performed in microgravity. We observe that, even in the absence of an external gravitational field, there is a transient period during which the average flowing velocities in the previously flowing regions are faster thus implying that transient weakening is occurring in microgravity.  In addition, the regions which were not previously flowing move significantly during reversal of the shear direction. This behaviour is not confined to the regions near the inner cylinder. In fact, particle motion can be observed across the \emph{entire width} of the shear cell, even in previously jammed regions. 

\begin{center}
 % [FIGURE 6 GOES HERE]
 [FIGURE \ref{fig:ShearRevProfs_4mm0g} GOES HERE]
\end{center}

\subsection{Quantifying the hysteresis}

To allow a direct comparison of the behaviour of the granular material under shear reversal in microgravity and on the ground we develop a method to quantify the extent of transient weakening.

First, we define the \emph{motor start time} as the moment when the mean angular velocity in the inner radial bin reaches ($1/5$) $\omega$ where $\omega$ is the inner cylinder angular velocity. The \emph{analysis start time} is then defined to be three seconds after the \textit{motor start time}.  As the mean particle velocity takes 1-2 seconds to reach a steady state, this gives more than long enough after motor start for 'steady' motion to be established. We also define the \emph{shear reversal time} to be the moment when the mean angular velocity of the inner radial bin passes through 0 rad s$^{-1}$.

To quantify the degree of transient weakening occuring in the granular system after shear reversal we first calculate the mean angular displacement travelled by the particles undergoing constant shearing for a strain of five particle diameters \ie during an inner cylinder displacement of 5 $d$\footnote{A strain of 5 $d$ is equivalent to an inner cylinder angular displacement of $\sim$0.2 rad for the 4 mm beads and $\sim$0.15 rad for the 3 mm beads.}. This is calculated as a function of distance from the inner cylinder. To calculate the mean angular displacement travelled by the particles the location of each particle at both the \emph{analysis start time} and after the time required for the inner cylinder to rotate 5 $d$ are found (these two times are indicated in \figs{detection}{ShearRevProfs_4mm0g} by the markers 1 and 2, respectively). The angular displacement of a particle is then given by the absolute difference between the two angular coordinates. The particles are then separated into the ten radial bins (\fig{RadialBins}) and the mean angular displacement, $S_\theta(r)$, of all particles within the radial bin is found.

Next the mean angular displacement is calculated, in exactly the same manner, for the particles undergoing the same strain immediately after shear reversal \ie starting from the shear reversal time and ending at the shear reversal time plus the time required for the inner cylinder to rotate 5 $d$ (these two times are indicated in \figs{detection}{ShearRevProfs_4mm0g} by the markers 3 and 4, respectively). 

Finally, the mean extra displacement, $L_e$, of particles during the transient state just after shear reversal is calculated, as a function of distance from the inner cylinder. This is given by the difference in the mean angular displacements travelled by the particles (for a strain of 5 $d$) just after shear reversal and during steady state shear. This is the same approach that was used to quantify the scale of the transient weakening in \cite{toiya04}.

\Fig{ExtraDisplacementEx} (a) shows the mean particle angular displacements for a strain of 5 $d$ under constant shearing and immediately after shear reversal for an example ground-based experiment with 4 mm particles and an inner cylinder angular velocity of 0.05 rad s$^{-1}$.  The extra displacement travelled by the particles just after shear reversal is shown in \fig{ExtraDisplacementEx} (b).  In the example ground-based experiment shown there is a mean extra displacement of $\sim$ 0.5 $d$ close to the inner cylinder just after shear reversal. Further from the inner cylinder, at a distance of $\gtrsim$7 $d$ there is no extra displacement of the particles after shear reversal.

\begin{center}
%  [FIGURE 7 GOES HERE]
 [FIGURE \ref{fig:ExtraDisplacementEx} GOES HERE]
\end{center}

%An identical analysis is performed looking at the bottom surface of the same experiment (\fig{ExtraDisplacementExBOTTOM}). In this example we can see that particles on the bottom surface near the inner cylinder have a negative extra displacement. This means that they move less distance just after shear reversal compared to during steady state shear. This agrees with our initial assessment based on the plots of angular velocity as a function of time that transient weakening does occur in our experiment on the ground, but only at the top (free) surface and not at the bottom surface where the granular material is crystallised and confined by the weight of the beads located above.

\subsection{Hysteresis in varying gravitational environments}

The mean angular displacement of particles undergoing a strain of five particle diameters is calculated for all experiments (constant shear rate and shear-reversal experiments in microgravity and on the ground) for which there are sufficient data. In other words, this is calculated for all experiments that neither end before the inner cylinder has moved 5 $d$, nor in which shear reversal occurs before the inner cylinder has moved 5 $d$. The mean angular displacement of particles undergoing a strain of five particle diameters just after shear reversal is also calculated for all shear reversal experiments for which there are sufficient data. The mean angular displacements, $\overline{S_\theta(r)}$, under constant shear and after shear reversal are then calculated for each experiment type. This allows a calculation to be made of the mean extra displacement of particles after shear reversal, $\overline{L_e(r)}$, for each experiment type.

Consider the extra displacement travelled by particles just after shear reversal for a strain of 5 $d$ (\fig{Cam2_ExtraDisp}). Close to the shearing surface the reversal of the shear direction causes a larger extra displacement in 1 $g$ than in microgravity. However, far from the shearing surface, there is a larger extra displacement of particles in microgravity compared to that in 1 $g$.  In fact, in 1 $g$ there is absolutely no movement in the regions far from the inner cylinder just after shear reversal. However, in microgravity we observe motion of the particles across the entire width of the shear cell just after shear reversal. We observe the same behaviour for both the 3 mm and the 4 mm glass beads.

This implies that the spatial extent of the transient weakening is enhanced in the microgravity environment meaning that the transient shear band after reversal of shear direction is wider in microgravity than on the ground.  However, close to the shearing surface the transient weakening may be reduced in the microgravity environment.

\begin{center}
  %[FIGURE 8 GOES HERE]
 [FIGURE \ref{fig:Cam2_ExtraDisp} GOES HERE]
\end{center}

\subsection{Experiment conclusions}

We have observed the occurence of hysteresis, in the form of transient weakening just after the reversal of shear direction, on the ground and in microgravity. By considering the mean extra displacement of particles just after shear reversal, \ie the difference in the mean angular displacements travelled by the particles (for a strain of 5 $d$) just after shear reversal and during steady state shear, we have been able to quantify and compare the extent of the hysteresis in the different gravitational regimes.

We have observed that, close to the shearing surface, the reversal of the shear direction causes a larger extra displacement in 1 $g$ than in microgravity. However, far from the shearing surface, there is a larger extra displacement of particles in microgravity compared to that in 1 $g$. This implies that the spatial extent of the transient weakening is enhanced in the microgravity environment meaning that the transient shear band after reversal of shear direction is wider in microgravity than on the ground.  However, close to the shearing surface the transient weakening may be reduced by the microgravity environment.

Perhaps compression of a granular material in the presence of an external gravitational field increases the coupling of the granular material to the walls and creates a stronger contact network between the particles than in microgravity. The breaking of a stronger contact network may then cause an enhanced transient weakening when an external gravitational acceleration is present, with respect to the transient weakening that may occur in microgravity. However, it would seem that, although the contact network may be weaker in microgravity the influence of any change in the contact network \ie during reversal of shear direction, is felt by the granular material over much larger distances.

\section{Implications for small body surfaces}\label{s:implications}

We have demonstrated that transient weakening of granular material occurs during shear reversal in microgravity as well as in the presence of an external gravitational field. {Of course, regolith is not expected to consist of identically sized, perfectly spherical particles. Indeed, asteroid regolith is more likely to be angular in shape and follow a power law size distribution \citep[\eg][]{miyamoto07}. Nonetheless, it has recently been found that in highly polydisperse granular systems the force chains are mainly captured by large particles. The shear strength of a non-cohesive granular material is, therefore, practically independent of particle size distribution \citep{voivret09}.   It has also been observed that the shear strength and force inhomogeneity are enhanced in a granular material with angular particles \citep{azema07}. Due to the increased importance of force chains between angular particles, we might then expect the effect of shear and shear reversal to be even more important for irregular shaped regolith. 

The hysteresis observed in our experiments after shear reversal may, therefore, be exploited to make a more efficient sampling mechanism for use on the surface of planets or small bodies. For example, to loosen the regolith a device could be used which reverses the direction of shear from time to time rather than using rotation in purely one direction. This may help to loosen and collect the required mass of regolith more rapidly or in a more power-efficient manner. Similarly, this method could be used to dig deeper into the regolith layer without consuming more power. 

However, it must be noted that the transient weakening observed in our experiment in the different gravitational environments suggests that, perhaps, compression of a granular material in the presence of an external gravitational field creates a stronger contact network between the particles than in microgravity. The breaking of a stronger contact network may then cause an enhanced transient weakening when an external gravitational acceleration is present, with respect to the transient weakening that may occur in microgravity.  This means that, in the low-gravity environment of an asteroid surface, while transient weakening will still be experienced by the granular material close to the location of shear reversal, it may not weaken the material as much as if the shear reversal occured in the presence of a stronger gravitational field.

However, it would seem that, although the contact network may be weaker in microgravity the influence of any change in the contact network \ie during reversal of shear direction, is felt by the granular material over much larger distances. In our experiment there was absolutely no motion at all at the outer edge of the shear cell during constant shearing in microgravity (except for the random motion due to the gravity fluctuations). However, just after shear reversal the particles at the very outer edge of the cylinder (\ie at $>$30 particle diameters from the shearing surface) moved! This may have very important implications for our interpretation of asteroid surfaces. If, for instance, a rubble pile asteroid has undergone shear forces in one particular direction for some period of time \eg due to tidal forces from planetary encounters \citep{bottke96} or systematic YORP spin up \citep{bottke06}, a very extended contact network may develop. Then, if an event was to occur which shears the surface in a different direction \eg an impact \citep{paolicchi02, holsapple02} the consequences could be incredibly long range. 

Perhaps, then, a small event, \eg a meteorite impact or a spacecraft landing, on one side of a small rubble pile asteroid could destabilise regolith on the other side of the asteroid causing a granular flow. This would occur, not necessarily via seismic shaking, but via the long range transmission of forces through the contact network. This could mean that the asteroid surfaces are even more unstable than previously imagined. However, this also indicates that the consequences of, \eg a meteorite impact or a spacecraft landing, may be very different depending on the impact angle and location, and depending on the prior history of the asteroid surface.

It would be interesting to perform further experiments investigating the effect of shear reversal on irregularly shaped particles. It would equally be interesting to investigate the influence of cohesion on transient weakening. This may be of particular importance for asteroids as it has been predicted that van der Waals cohesion should be a dominant force between regolith particles on asteroid surfaces \citep{scheeres10}.

\section*{Acknowledgments}
Thanks to The Open University, Thales Alenia Space, the UK Science and Technologies Facilities Coucil, the Royal Astronomical Society and the French National Programme for Planetology for providing financial support. Thank you also to ESA ÔFly your ThesisÕ for giving us the opportunity to be part of the 51st ESA microgravity research campaign, and for the financial support. Finally, thanks to the workshop of the Department of Physical Sciences at The Open University for constructing our experimental hardware. Part of this study was done as part of the International Team (\#202) sponsored by the International Space Science Institute (ISSI), Bern, Switzerland.

\bibliographystyle{mn2e_v2} % bibliography style
\bibliography{/Users/naomimurdoch/Documents/BIBFILE/murdoch.bib} % References file

\section*{FIGURES}

\begin{figure*} 
\centering
\subfigure[]{
 \includegraphics[height=0.7\columnwidth]{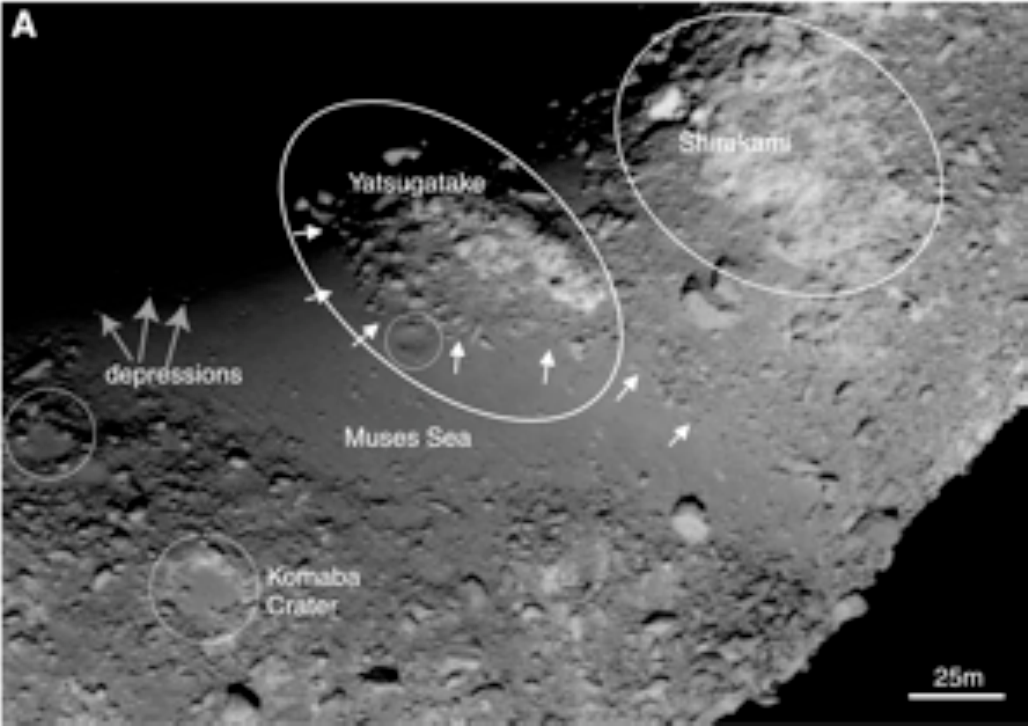}
   \label{fig:MusesSeaA}
}
\subfigure[]{
 \includegraphics[height=0.7\columnwidth]{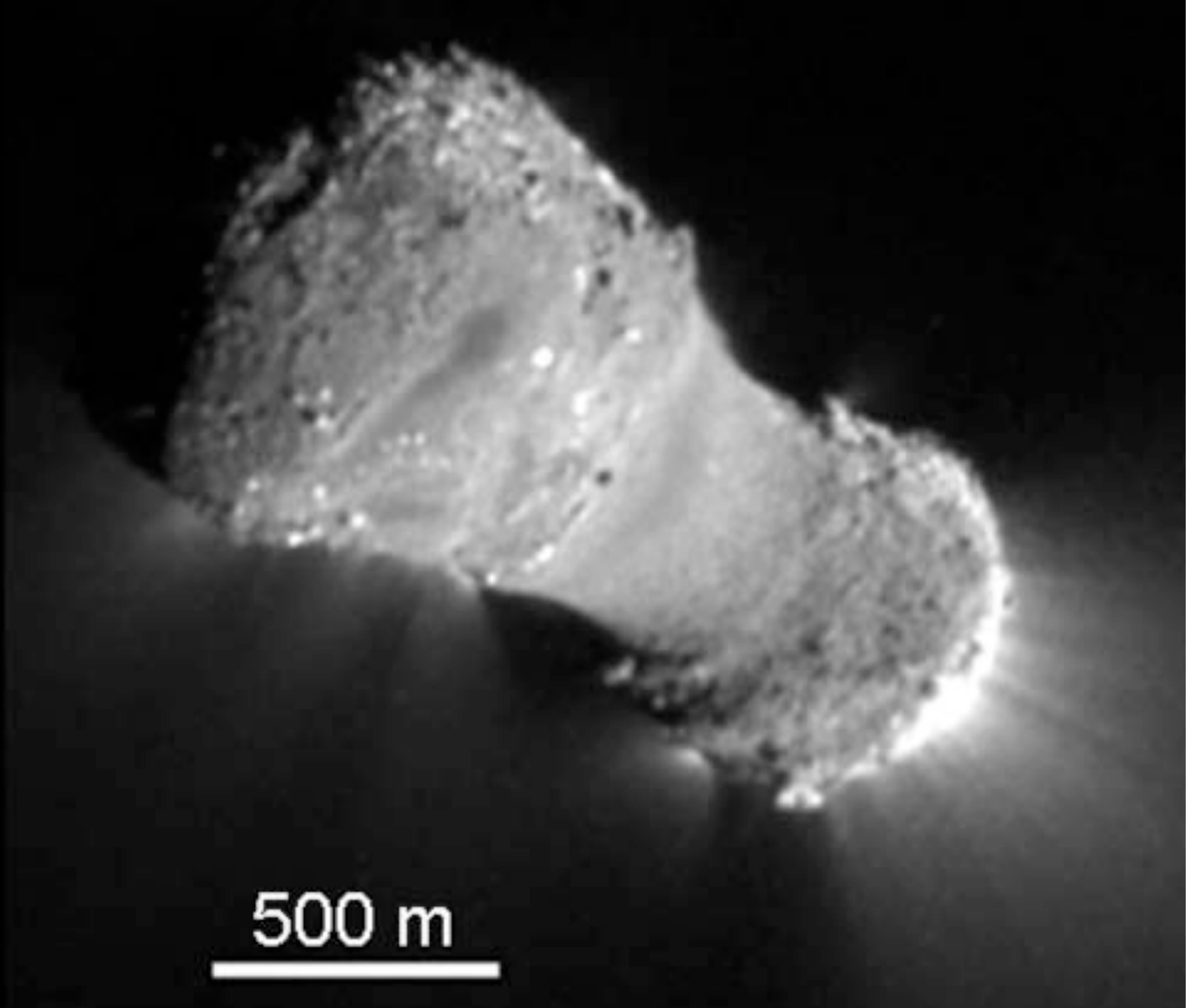}
  \label{fig:HARTLEY2}
}
\caption{(a) Evidence of downslope movement on (25143) Itokawa. The small white arrows in the Muses Sea region indicate the thin, boulder-rich layer similar to landslide deposits. Image taken by the Hayabusa spacecraft; figure from \protect \citep{saito06}. (b) Image of comet 103P/Hartley 2 clearly showing segregation of grains on the surface. \protect \cite{ahearn11} suggest that the smooth shape of the ``waist'' region connecting the two lobes might indicate material collecting in a gravitational low. Image taken by the EPOXI spacecaft; figure from \protect \citep{ahearn11}.}
\label{fig:RegolithSeg}
\end{figure*}

\begin{figure*} 
\centering
 \includegraphics[width=1.5\columnwidth]{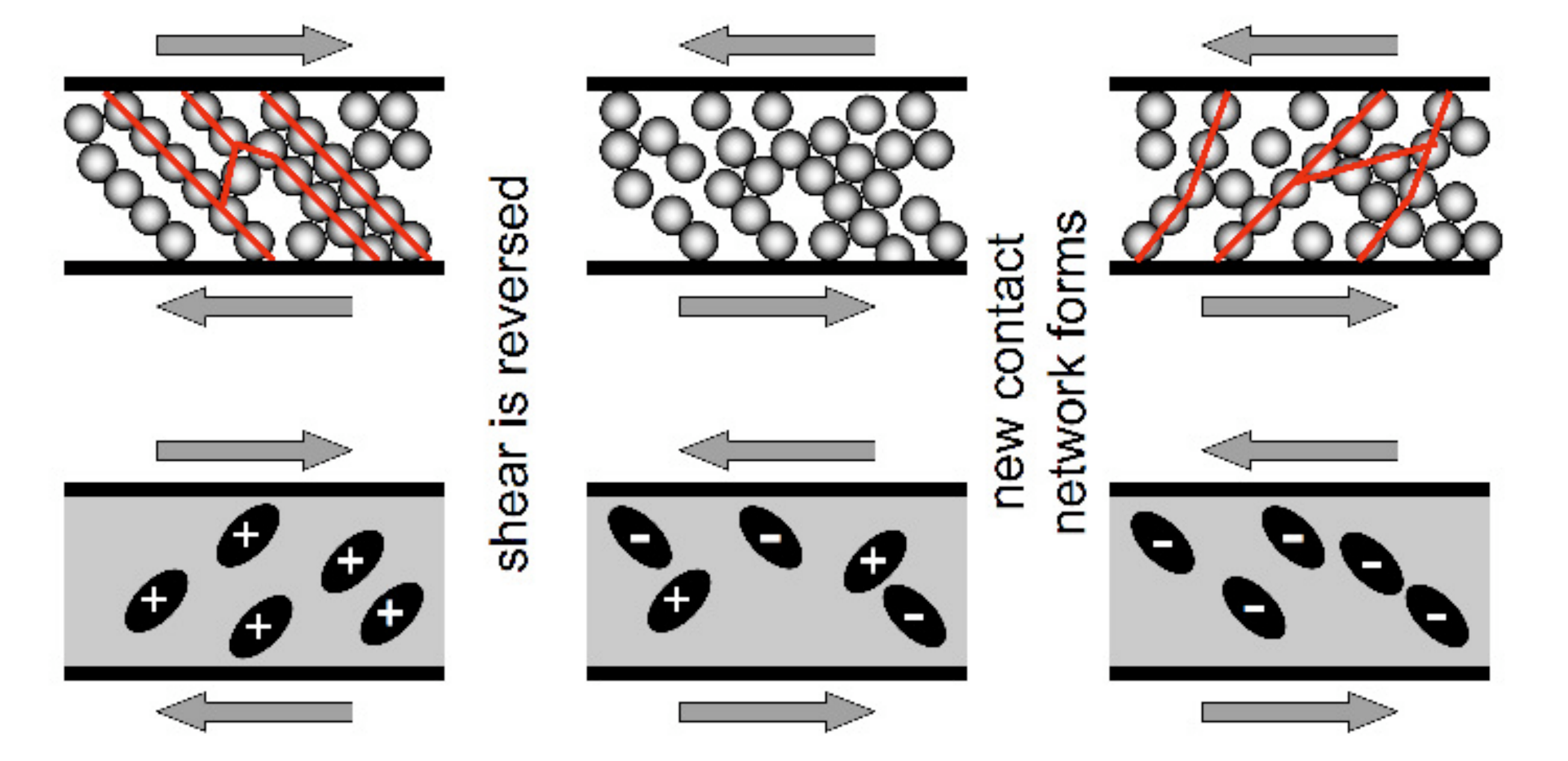}
 \caption{Schematic of shear reversal experiments. The breaking and re-forming of a contact network through shear reversal in dry granular matter is illustrated schematically. The arrows indicate the relative movement of the cylinder walls and the lines through the particles in the upper diagrams indicate the stress transmission. The positive and negative signs show the principal direction of the stress transmission. When shear is started opposite to the prior shear direction, transiently the material compacts and is easy to shear. Image from \protect \cite{falk08}.}
  \label{fig:shear_reversal}
\end{figure*}

\begin{figure*}
\centering
\includegraphics[width=0.7\columnwidth]{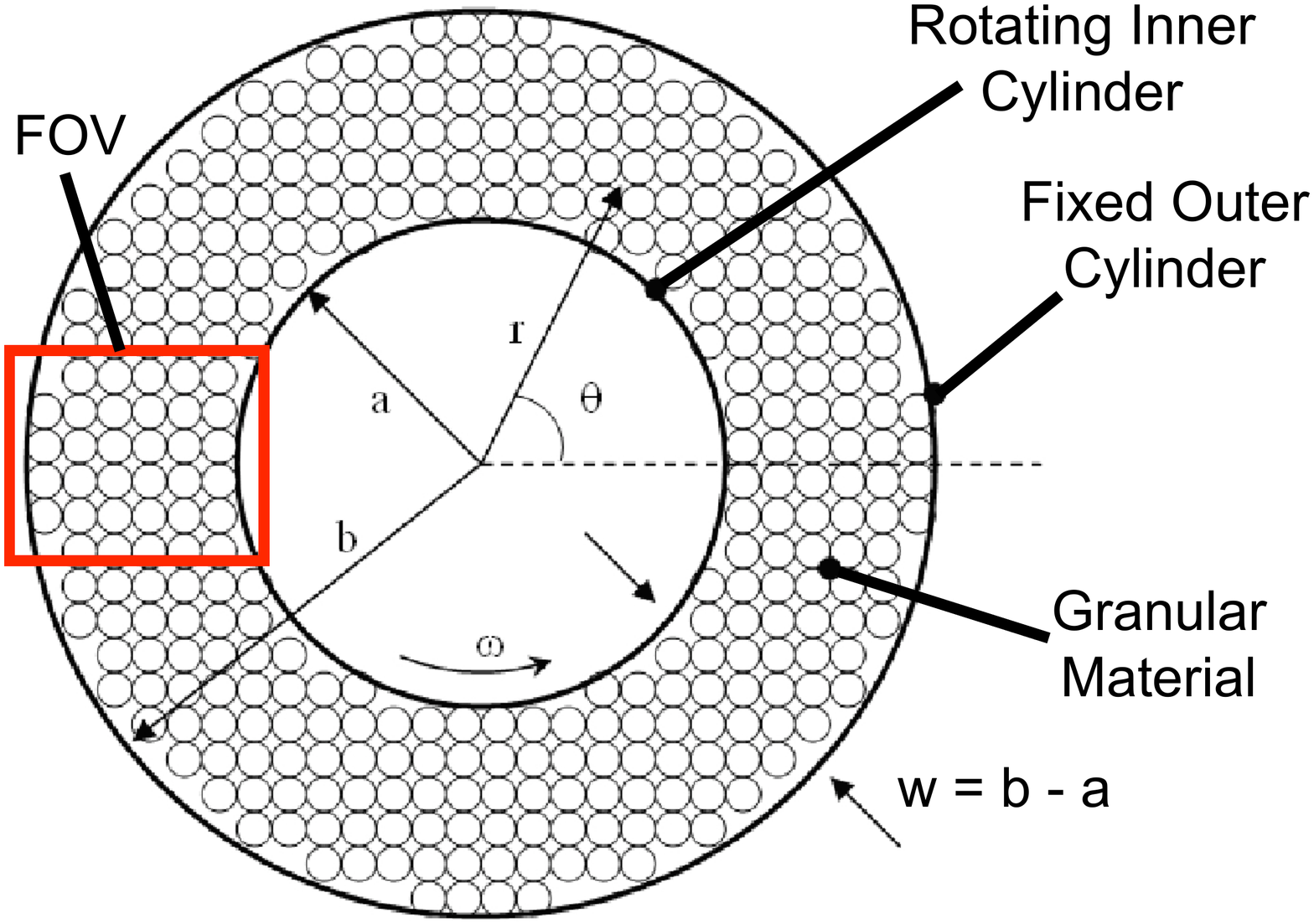}
\caption{The Taylor-Couette Geometry ($a$ = inner cylinder radius, $b$ = outer cylinder radius, $w$ = width of shear region, $r$ = radial distance, $\theta$ = angular distance, $\omega$ = inner cylinder angular velocity). The camera field of view (FOV) is also shown. Image adapted from \protect  \cite{toiya06}.}
\label{fig:TaylorCouette}
\end{figure*}

 \begin{figure*} 
\centering
 \includegraphics[width=1\columnwidth]{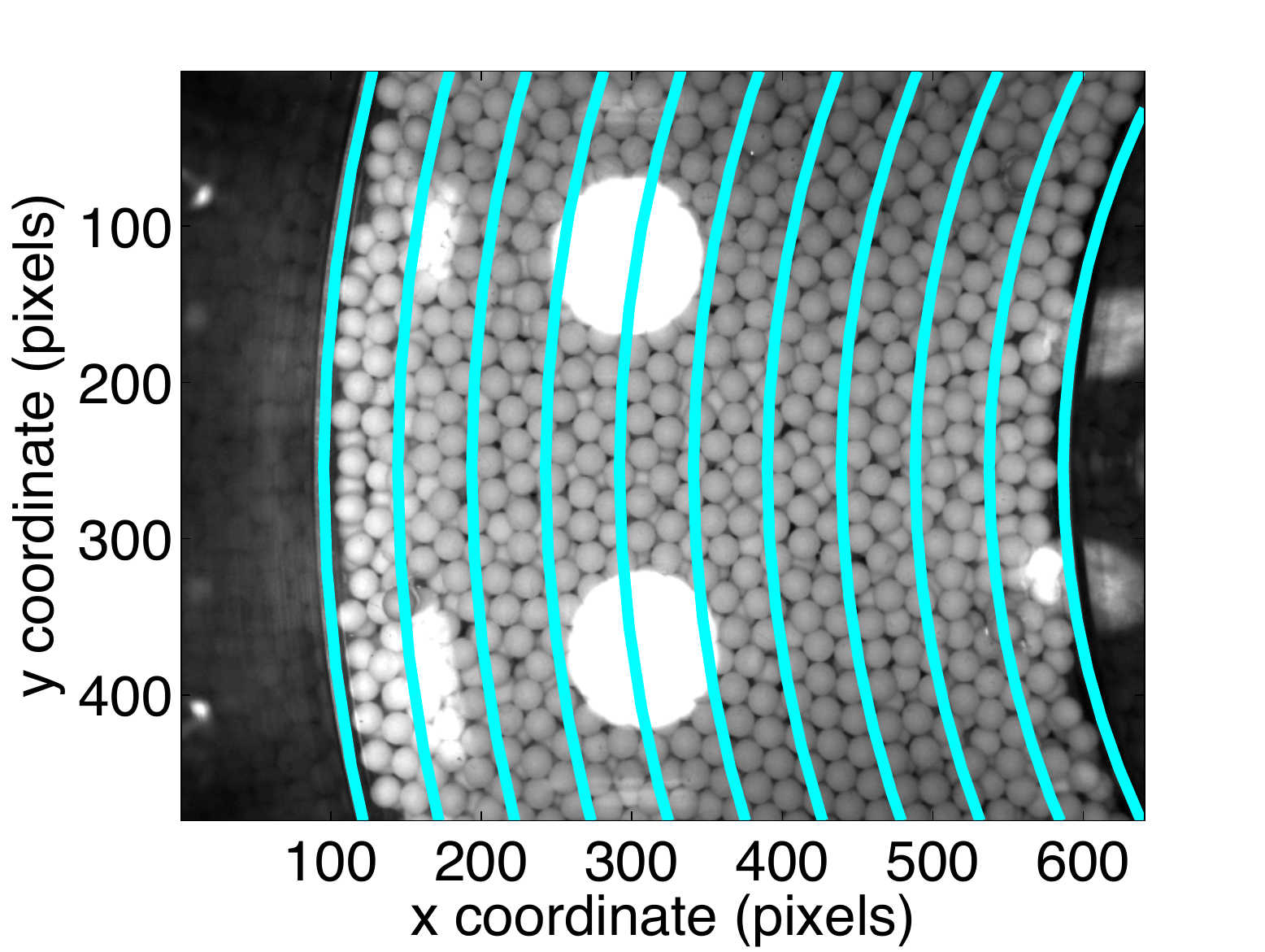}
 \caption{Example image showing the radial bins used during the analysis. One image with the boundaries of the ten radial bins plotted over the top. The particles are split into these radial bins based on their radial position. Radial bin 1 is closest to the inner, rotating, cylinder (\ie furthest to the right in this image). The two large bright white spots are the reflections of the lamps on the movable and transparent disk that is used to confine the granular material.}
 \label{fig:RadialBins}
 \end{figure*}
% above plots made with /extern2/murdoch/ASTEX/ASTEX_SCRIPTS/particle_tracking_scripts/CodesForTracking/PlotParticlesAndRadialBins.m

% profiles made with ExtraDisplacement.m
\begin{figure*} 
\centering
\subfigure[]{
 \includegraphics[width=1\columnwidth]{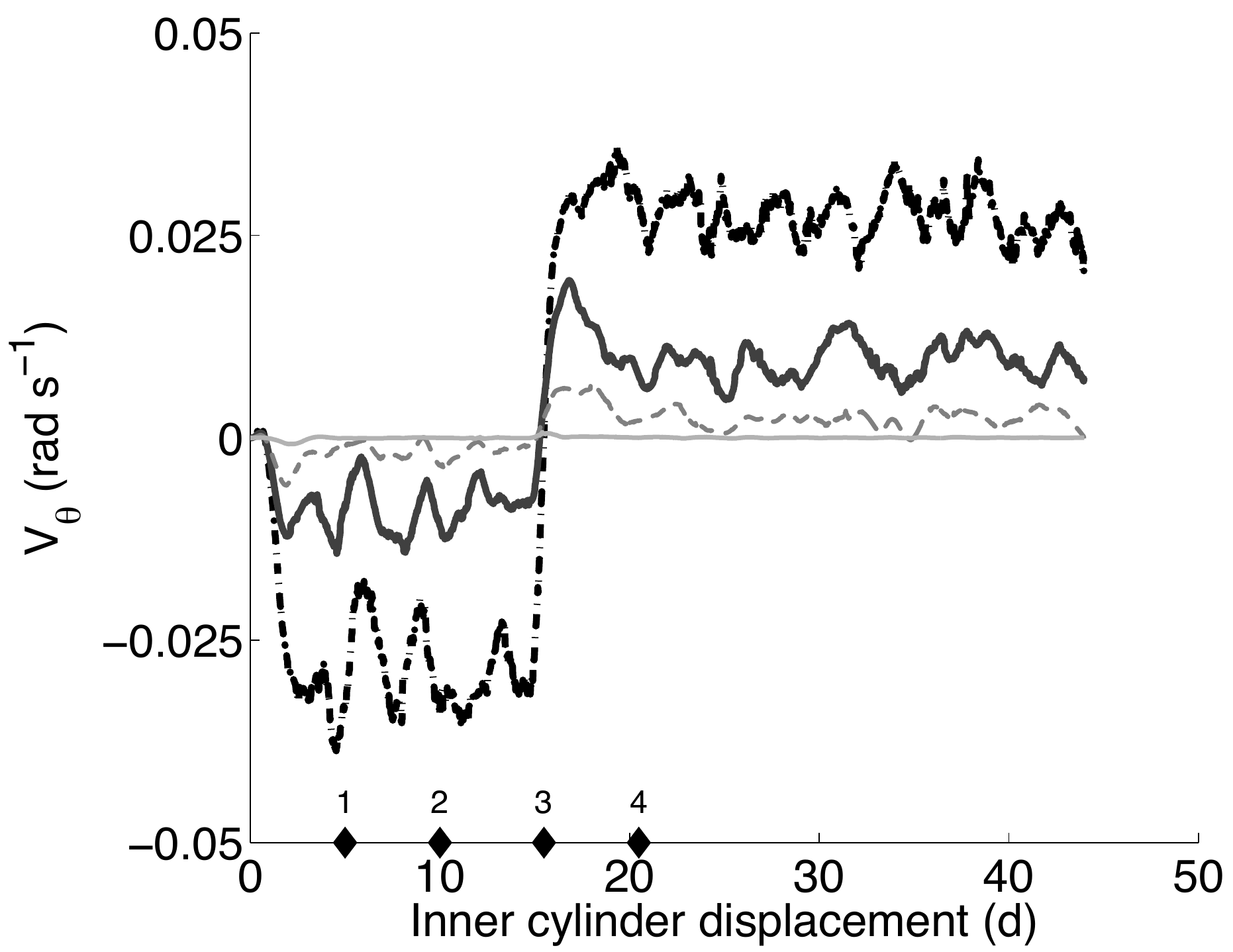}
 }
 \subfigure[]{
  \includegraphics[width=1\columnwidth]{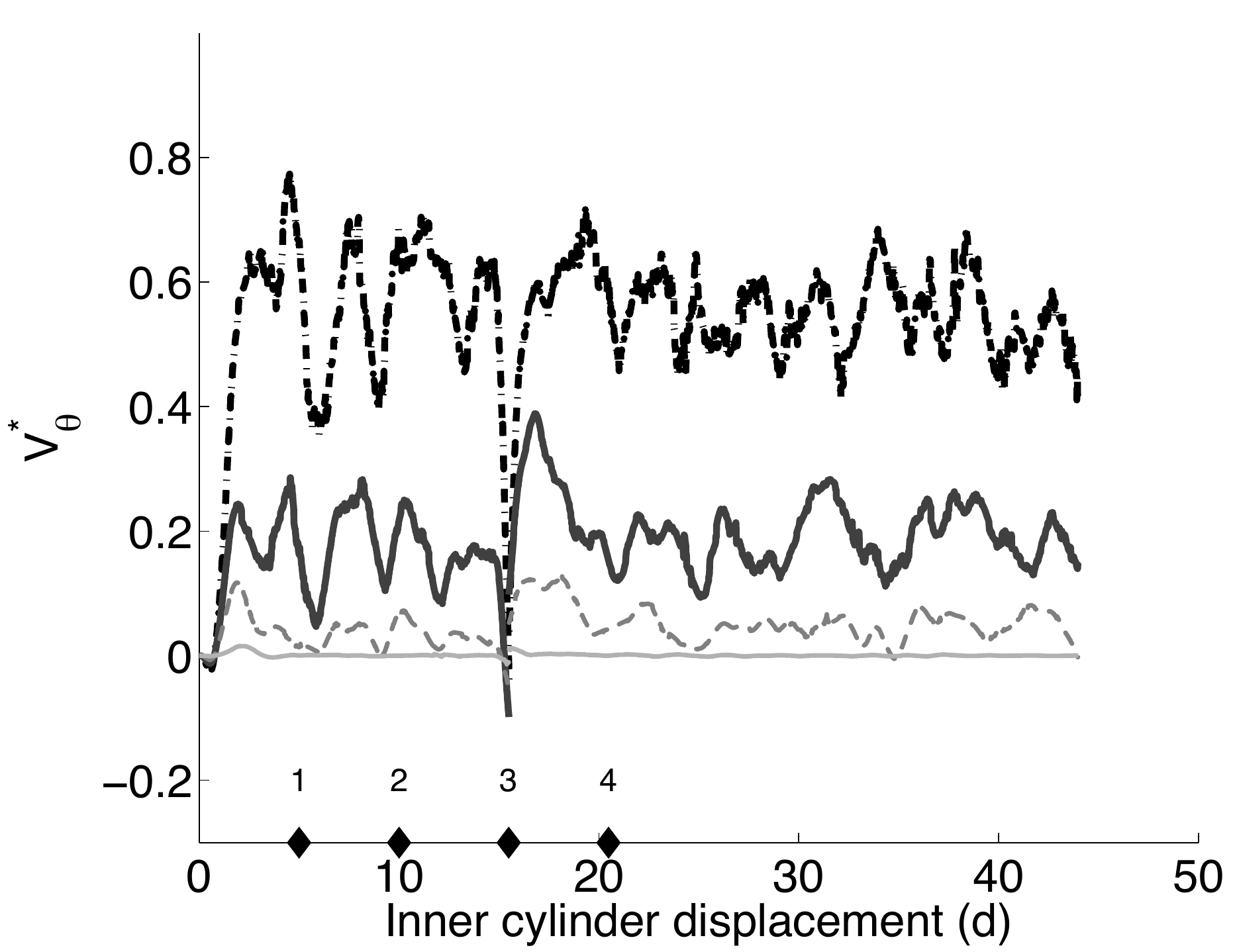}
}
%when insert put main figure width to 1.5
%\put(-170,37) {\includegraphics[width=0.6\columnwidth]{Camera2_GroundA11_ProfileWithTimeMarkers}}
\caption{Shear reversal experiment in 1 $g$.  (a) $V_\theta$ and (b) $V^*_\theta$  (see \eqn{VstarTheta}) for four radial bins versus inner cylinder displacement in particle diameters for shear reversal experiments performed on the ground. The inner cylinder angular velocity was 0.05 rad s$^{-1}$. The particles are divided into ten radial bins. The order, from closest to the shearing surface to furthest away: bin 1 - dashed-dotted broad line, bin 2 - solid broad line, bin 3 - narrow grey dashed line, bin 10 - narrow grey solid line.  The data for bins 4 to 9 all lie within the results for bins 3 and 10. The numbered markers indicate (1) the \emph{analysis start time}, (2) \emph{analysis start time} plus the time required for the inner cylinder to rotate 5 $d$, (3) \emph{shear reversal time}  and (4) \emph{shear reversal time} plus the time required for the inner cylinder to rotate 5 $d$.}
\label{fig:detection}
 \end{figure*}

\begin{figure*} 
\centering
 \includegraphics[width=1.5\columnwidth]{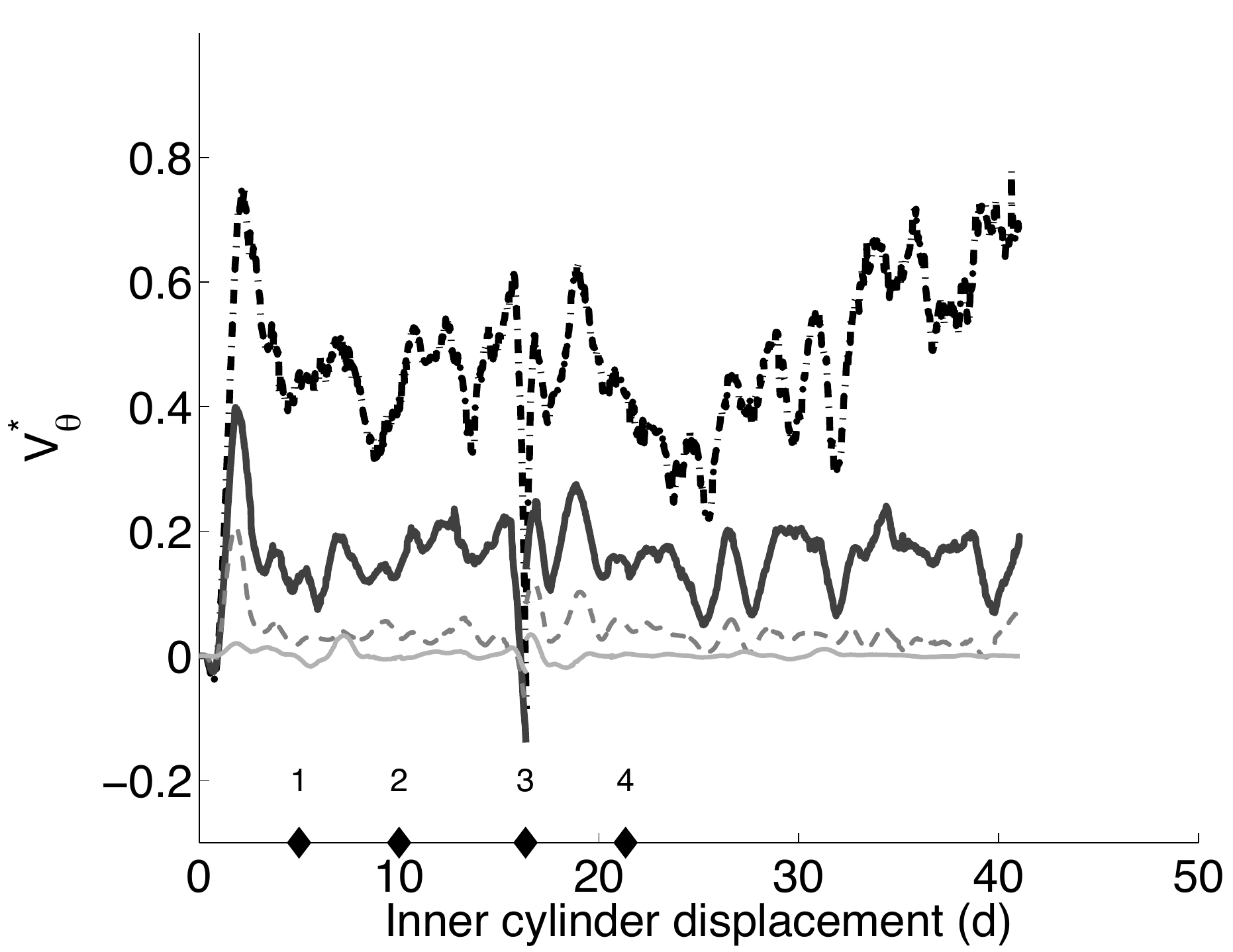}
\caption{Top surface during a shear reversal experiment with 4 mm beads in microgravity. $V^*_\theta$ (see \eqn{VstarTheta}) for four radial bins versus inner cylinder displacement in particle diameters for shear reversal experiments performed in microgravity. The inner cylinder angular velocity was 0.05 rad s$^{-1}$. The lines and markers are the same as in \fig{detection}.}
\label{fig:ShearRevProfs_4mm0g}
 \end{figure*}

 % plots made with ExtraDisplacement.m (on harddisk)

 \begin{figure*} 
\centering
 \subfigure[]{ 
  \includegraphics[width=1\columnwidth]{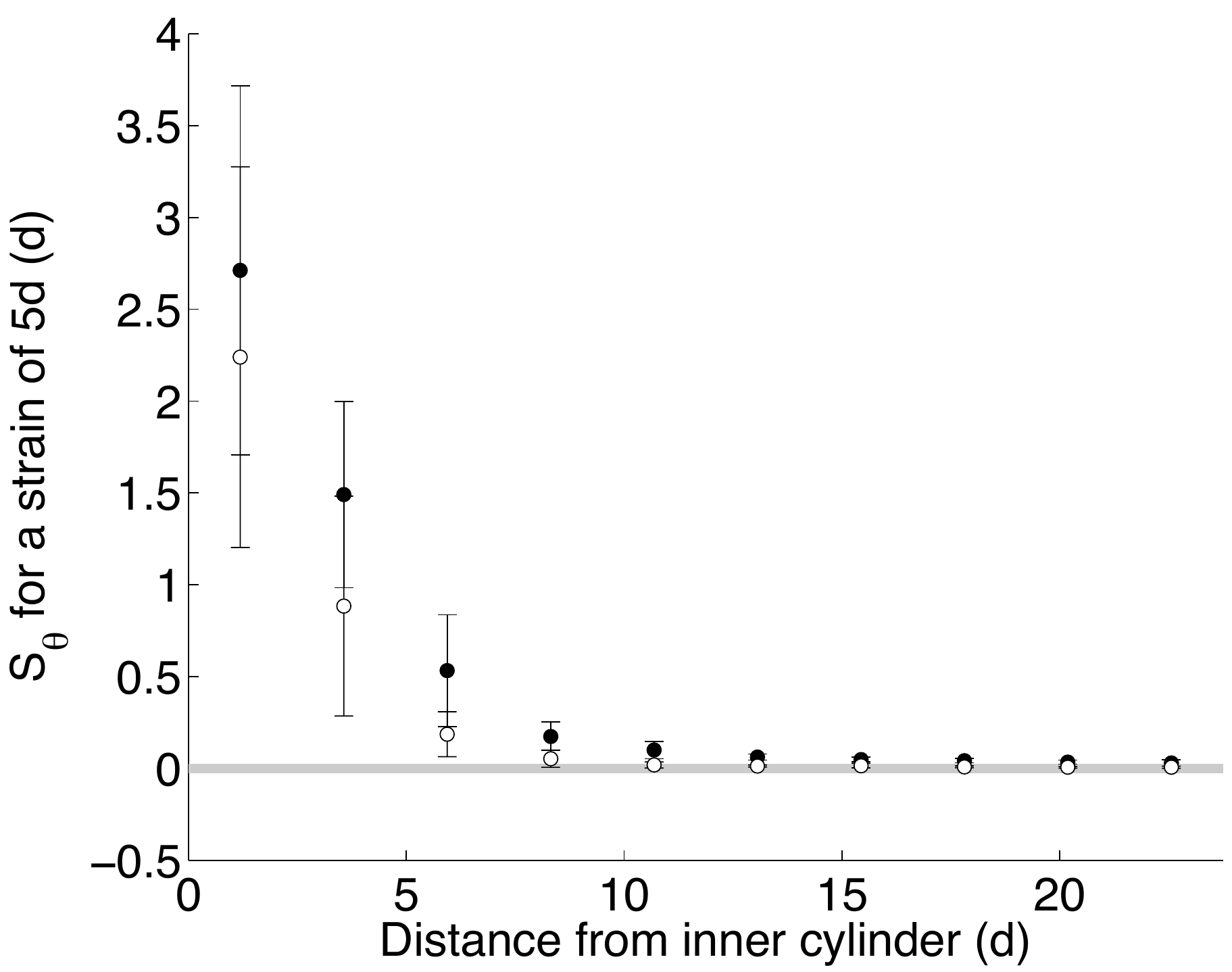}
  }
   \subfigure[]{ 
  \includegraphics[width=1\columnwidth]{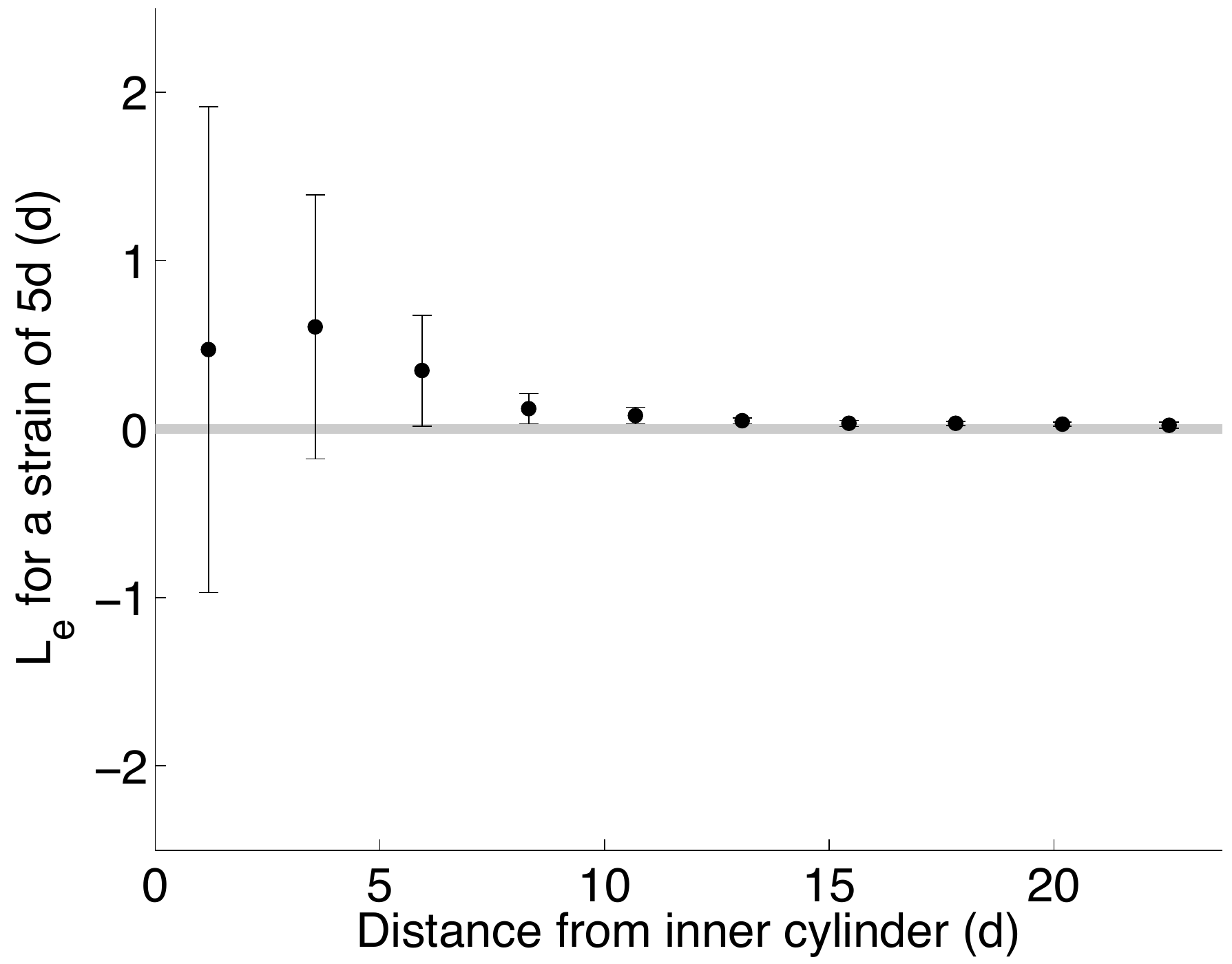}
  }
  \caption{The extra displacement of particles after shear reversal. (a) The mean particle angular displacements, $S_\theta(r)$, on the top surface for a strain of 5 $d$ under constant shearing (open markers) and immediately after shear reversal (solid markers) for an example ground-based experiment with 4 mm particles and an inner cylinder angular velocity of 0.05 rad s$^{-1}$. (b) The extra displacement, $L_e(r)$, of the particles just after shear reversal. The error bars in (a) represent the standard deviation of the mean particle angular displacement. The standard deviation in the constant shearing mean displacement and the shear reversal mean displacement are combined in quadratic to give the scatter in $L_e(r)$, represented by the error bars in (b).} 
  \label{fig:ExtraDisplacementEx}
 \end{figure*}
 
% CompareExtraDisplacements.m (with set running from 1:1)
% line 114
%    for group=size(MeanConstDisp,1)-1:-1:1 for 4mm
%    for group=size(MeanConstDisp,1):-1:1 for 3mm)

 \begin{figure*} 
\centering
 \subfigure[]{  
 \includegraphics[width=1\columnwidth]{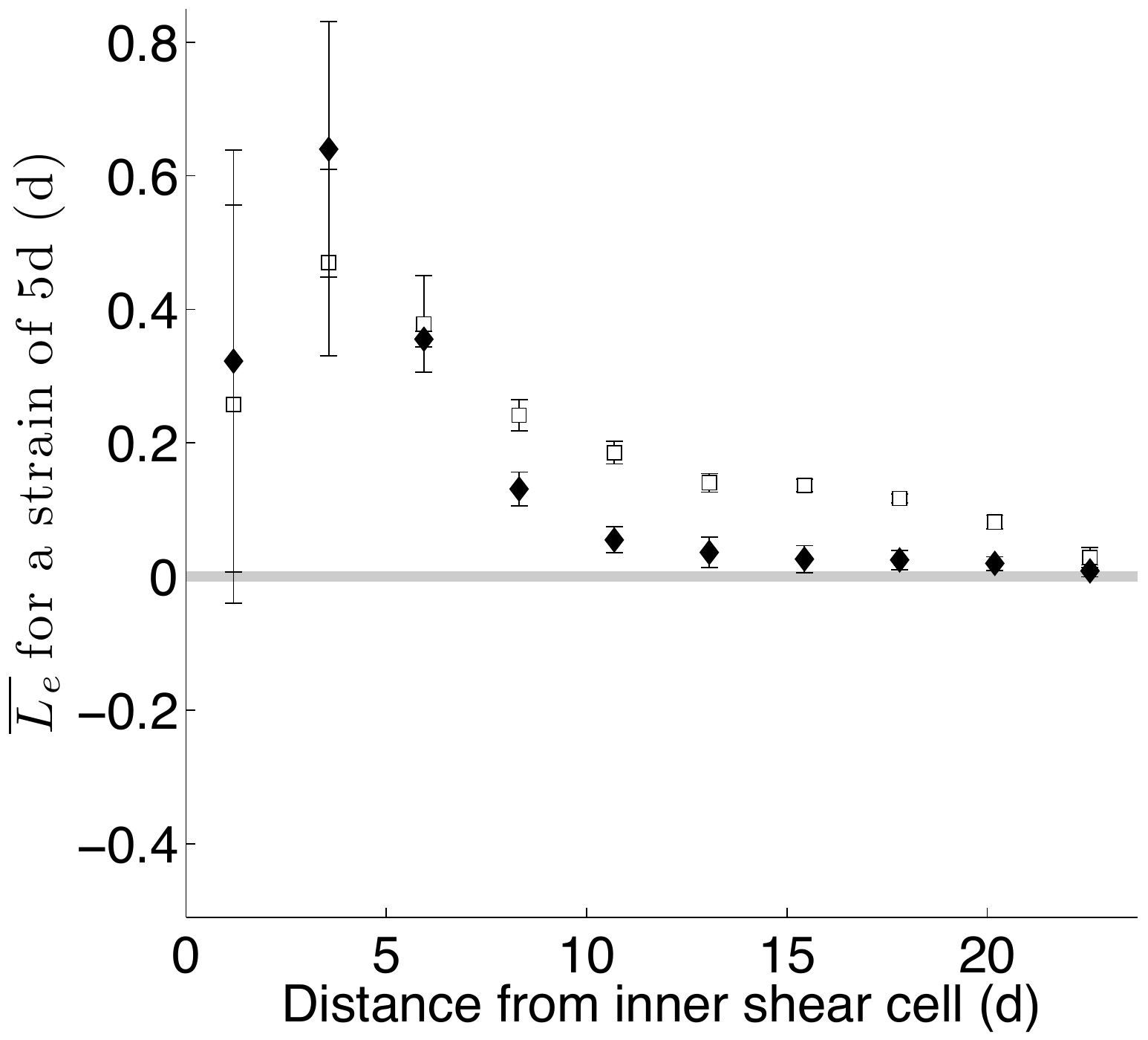}
  }
   \subfigure[]{ 
  \includegraphics[width=1\columnwidth]{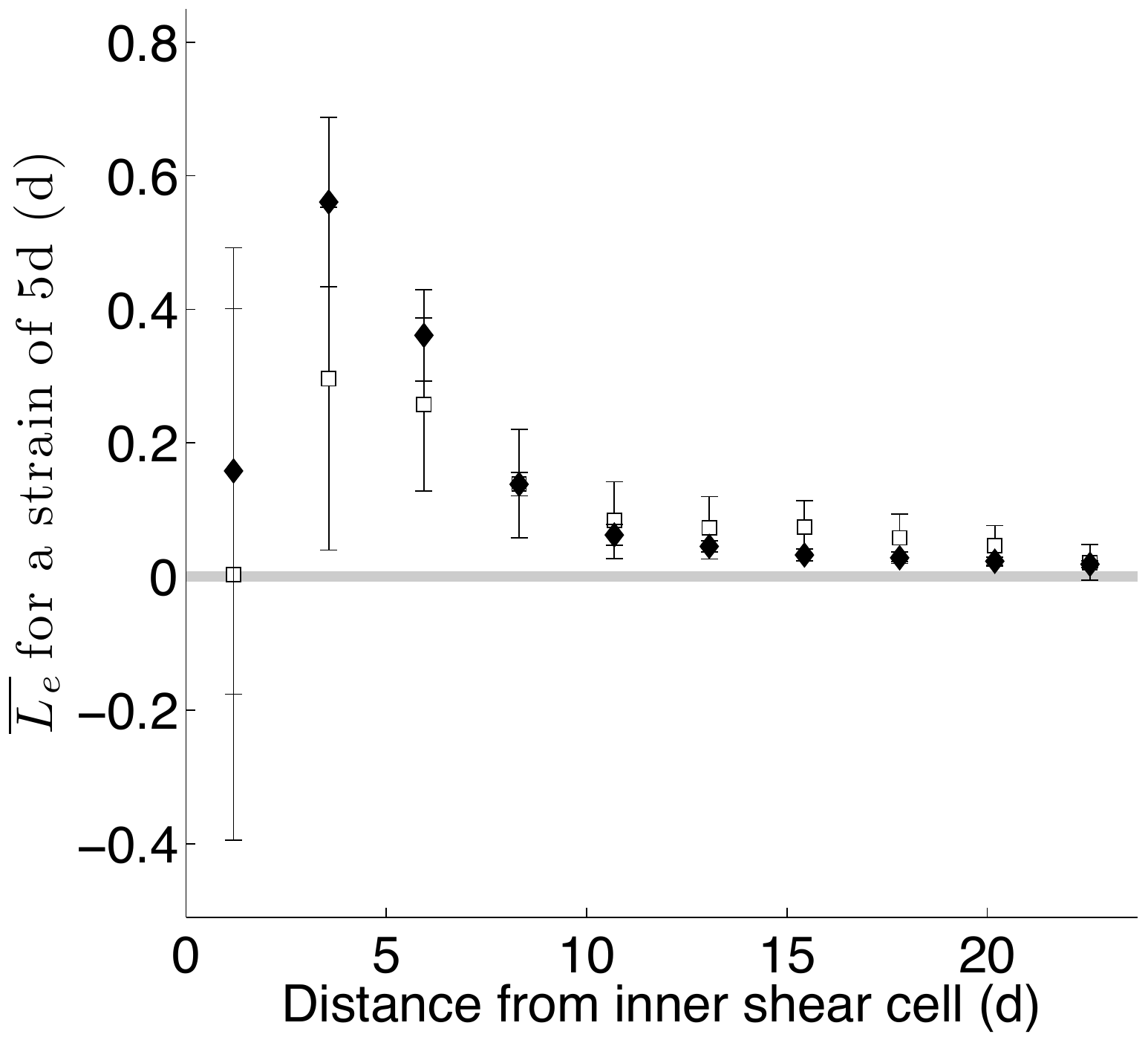}
  }
   \subfigure[]{  
 \includegraphics[width=1\columnwidth]{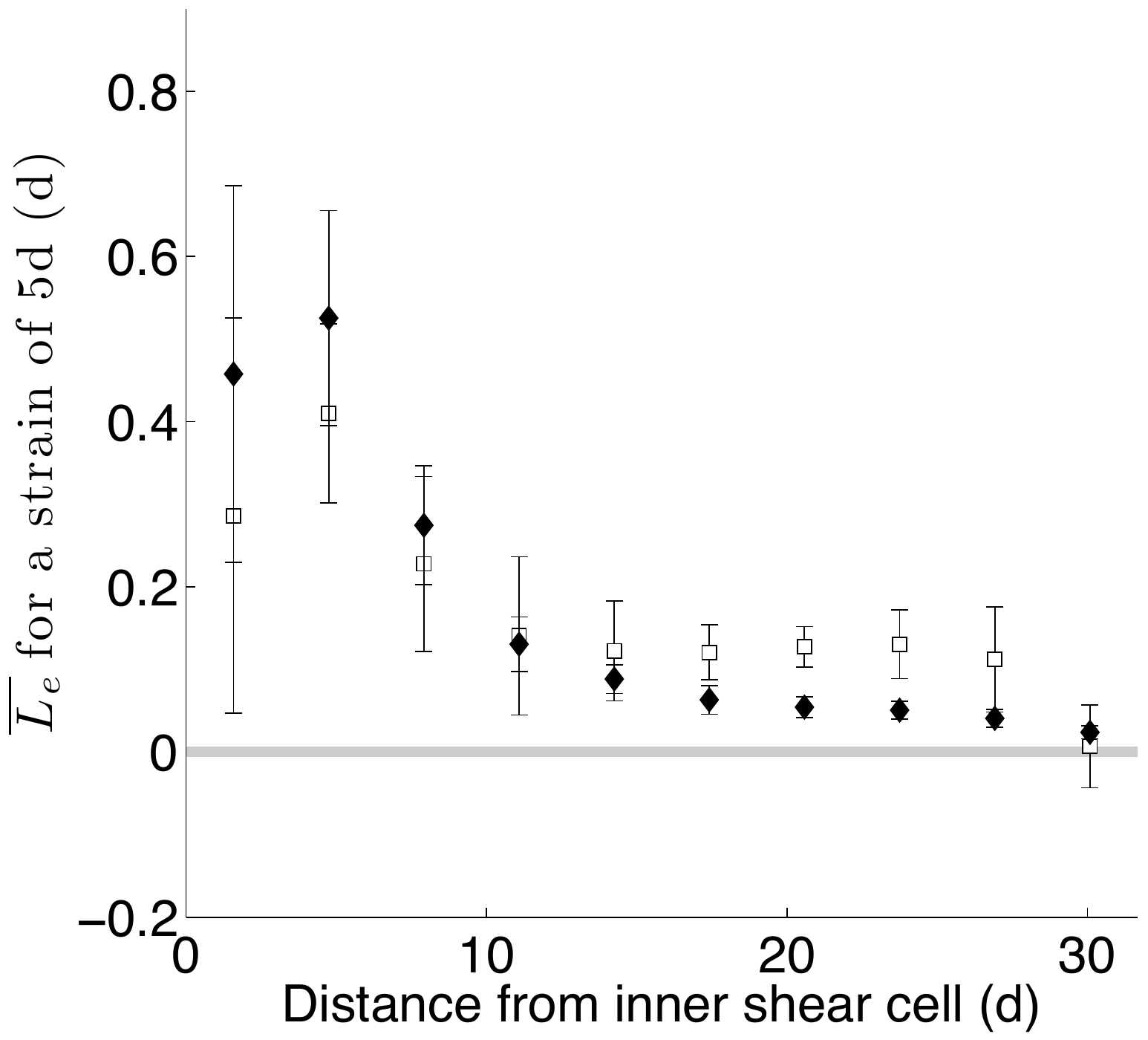}
  }
   \subfigure[]{ 
  \includegraphics[width=1\columnwidth]{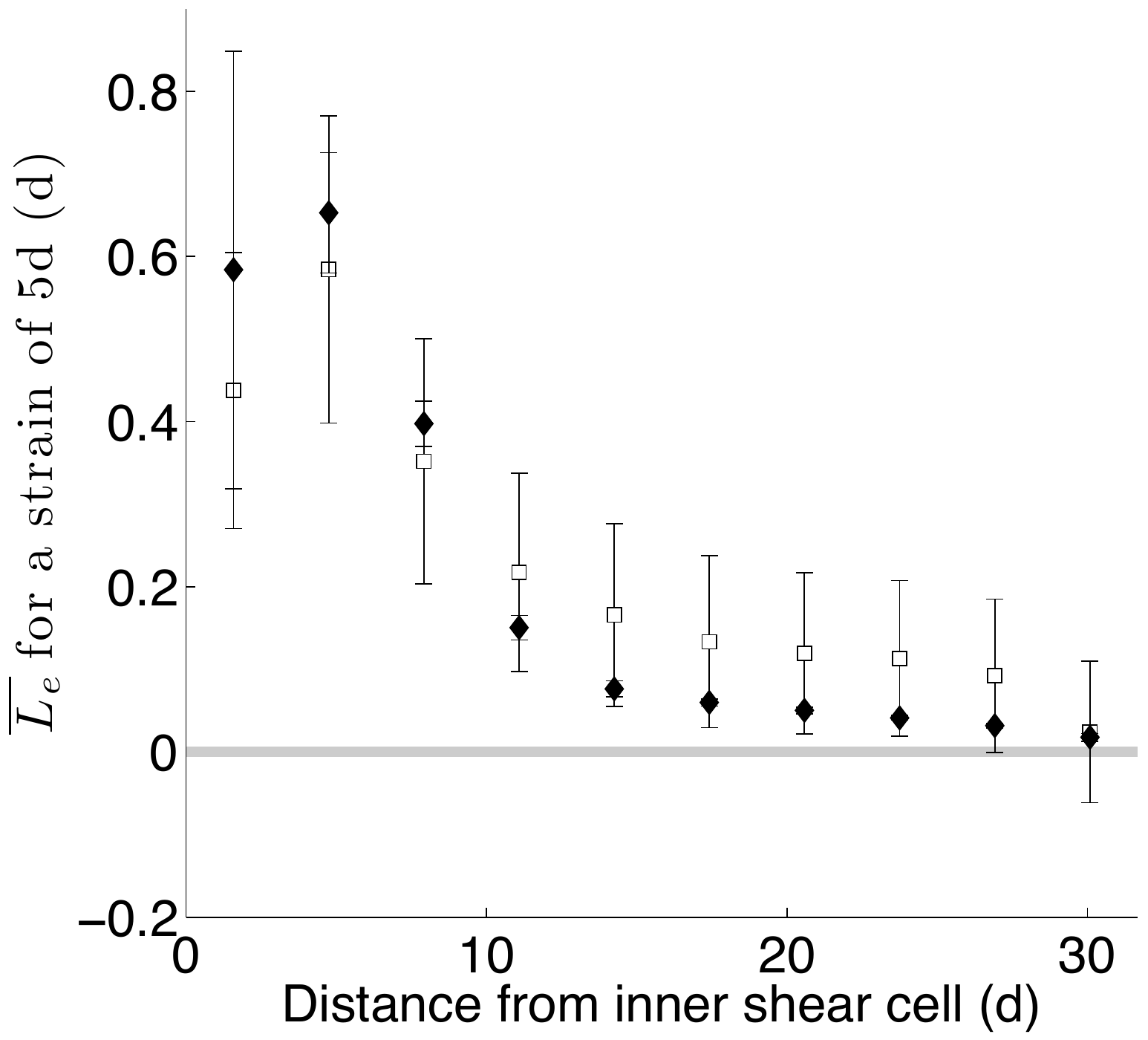}
  }
  \caption{Extra displacement of particles on the top surface after shear reversal. $\overline{L_e}$ as a function of distance from the inner cylinder for 4 mm (top) and 3 mm (bottom) particles. The inner cylinder angular velocity in the examples is 0.025 rad s$^{-1}$ (left) and 0.05 rad s$^{-1}$ (right).  $\overline{L_e(r)}$ is given for each experiment type: ground-based (black diamonds), microgravity (open squares). The error bars represent the scatter of $L_e(r)$ between different experiments of the same type.}
 \label{fig:Cam2_ExtraDisp}
\end{figure*}

\label{lastpage}

\end{document}